\documentstyle[12pt]{article}
\newcommand\beq{\begin{equation}}
\newcommand\eeq{\end{equation}}
\newcommand\bea{\begin{eqnarray}}
\newcommand\eea{\end{eqnarray}}

\begin{document}
\vspace{-2.0cm}
\bigskip

\begin{center} 
{\Large \bf  Coherent States with SU(2) and SU(3) Charges} 
\end{center} 
\vskip .8 true cm

\begin{center} 
{\bf Manu Mathur} \footnote{manu@bose.res.in} and 
{\bf Samir K. Paul}  \footnote{smr@bose.res.in}  
\vskip 0.5 true cm

S. N. Bose National Centre for Basic Sciences \\ 
JD Block, Sector III, Salt Lake City, Calcutta 700 098, India

\vskip .1 true cm

\end{center} 
\bigskip

\centerline{\bf Abstract}

\noindent We define coherent states carrying  SU(2) charge 
by exploiting Schwinger boson representation of SU(2) Lie 
algebra. These coherent states satisfy continuity property and provide 
resolution of identity on $S^{3}$.  We further generalize these 
techniques to construct the corresponding SU(3) charge coherent states. 
The SU(N) extension is also discussed. 
\vskip .4 true cm
\noindent PACS: ~03.65.-w, 02.20. Sv, 02.20.-a  
\vskip .4 true cm

\section{\bf Introduction}

The concept of coherent states was introduced by Schr\"odinger 
\cite{schrodinger} in the context of a harmonic oscillator. 
These harmonic oscillator 
coherent states, also called canonical coherent states, have been 
useful and studied extensively in physics \cite{klauder}. 
The next most important coherent states are spin coherent states or 
SU(2) coherent states which are 
associated with angular momentum or SU(2) group. Like canonical coherent 
states, they too have found wide applications in different branches of 
physics such as  quantum optics, statistical mechanics, nuclear physics 
and condensed matter physics  \cite{klauder}. It is 
known that these spin coherent states can also be constructed 
using harmonic oscillators by exploiting either Holstein Primakov 
or Schwinger boson representation of SU(2) Lie algebra 
\cite{schwinger,radcliffe,eriksson}. This harmonic oscillator 
formulation of spin 
coherent states is appealing because it is simple and analogous to 
canonical coherent state construction. Further, it bypasses the action 
of group elements \cite{perelomov} to get the spin coherent states. Motivated 
by the resulting simplifications, we recently generalized  this 
harmonic oscillator formulation of coherent states to SU(N) group 
\cite{manu}. In this work, we further exploit the above 
ideas to construct SU(2) and SU(3) charge coherent states defined on 
$S^3$ and $S^5$ respectively. The coherent states carrying 
SU(2) and SU(3) (non-abelian) charges  in two and three mode Fock spaces 
have been discussed in the past \cite{eriksson,bdr,fan,seok}. However 
they are defined on full complex planes  and are different from the 
SU(2) and SU(3) charge coherent states discussed in this paper which are defined 
on the compact manifolds $S^{3}$ and $S^5$ respectively.  We will further elaborate 
on these differences as we proceed (section 2.1 and section 3).  
 
\noindent The plan of the paper is as follows: We start with a brief description of 
harmonic oscillator coherent states. The coherent states discussed later will have their 
roots in this simple construction. In section 2, using 
two harmonic oscillators, we exploit Schwinger boson representation to construct SU(2) 
coherent states. This construction is known and is contained in 
\cite{schwinger}. 
However, we have included this section to make the presentation self contained. 
In section 3 we define SU(2) charge  coherent states which satisfy resolution of 
identity over the SU(2) group manifold $S^{3}$. 
In section 4 we generalize these ideas to SU(3) group. In section 5 we give 
SU(N) construction. 
  
The harmonic oscillator coherent states are defined as:  
\bea 
\vert z \rangle & = & \sum_{n=0}^{\infty} \frac {(z a^{\dagger})^{n}}{n!} \vert 0> \nonumber \\  
& = & \sum_{n=0}^{\infty} \frac {z^{n}}{\sqrt{n!}} \vert n>  
\label{cs1} 
\eea 
These coherent states are associated with Heisenberg-Weyl group whose Lie algebra is given by: 
\bea
[a,a^{\dagger}] = {\cal I}, ~~[a,{\cal I}]=0,~~ [a^{\dagger},{\cal I}]=0.  
\label{hw}
\eea
In (\ref{hw}), ${\cal I}$ is the identity operator. 
The manifold corresponding to the Heisenberg-Weyl 
group is the complex z plane. The coherent states in (\ref{cs1}) 
are  analytic  over this group manifold ${\cal M}$  and 
satisfy the resolution of identity: 
\bea 
\int_{\cal M}  d\mu(z) \vert z \rangle_{_{\infty}}~ 
{}_{_{\infty}}\langle z \vert 
\equiv {\cal I}.  
\label{roihw}  
\eea
\noindent In (\ref{roihw}), $d\mu{(z)} \equiv exp(-|{z}|^{2})~ dz d\bar{z}$ 
is the  measure over ${\cal M}$. We now give the construction of SU(2) coherent 
state which is similar to (\ref{cs1}). 

\section{\bf SU(2) Coherent States}

The SU(2) group involves 3 angular momentum generators, $J_1, J_2$ and 
$J_3$ and the Lie algebra is: 
\bea
[J^{a},J^{b}] = i~\epsilon_{abc}~ J^{c}, ~~~ a,b,c =1,2,3.   
\label{am}  
\eea 
\noindent The SU(2) Casimir operator is $\vec{J}.\vec{J}$ with eigenvalue 
j(j+1) where j is integer or half integer spin. 
The angular momentum algebra in (\ref{am}) can be realized in terms of
a doublet of harmonic oscillator annihilation creation  operators
$\vec{a} \equiv (a_1, a_2)$ and $ \vec{a}^{\dagger} \equiv (a^{\dagger}_1,
a^{\dagger}_2)$ respectively \cite{schwinger}. The number operators are 
$\hat{N}_1 \equiv a^{\dagger}_{1}a_{1}$ and $\hat{N}_2 \equiv 
a^{\dagger}_{2}a_{2}$ They satisfy the bosonic commutation relations: 
\bea
[a_i,a^{\dagger}_j] = \delta_{ij},~~~  i,j =1,2. 
\label{ho} 
\eea 
\noindent The vacuum state is $\vert 0,0 >$ and 
the number operator basis is written as $\vert n_1,n_2 >$. It 
satisfies: 
\bea
\hat{N}_1 ~ \vert n_1,n_2 >  =  n_1~\vert n_1,n_2 >, ~~~ 
\hat{N}_2 ~ \vert n_1,n_2 >   =  n_2~ \vert n_1,n_2 >. 
\label{no}
\eea 

\noindent We can now define the angular momentum operators in (\ref{am}) 
as: 
\beq
J^a ~\equiv  {1 \over 2} ~a^{\dagger}_i ~(\sigma^a)_{ij} ~a_j ~,
\label{sch}
\eeq
\noindent where $\sigma^a$ denote the Pauli matrices.  
It is easy to check that the operators in (\ref{sch}) satisfy (\ref{am}). 
Further, as they involve one creation and one annihilation operator, 
the Casimir $({\cal C})$ is: 
\bea 
{\cal C}  = \hat{N}_1  +   \hat{N}_2.  
\label{cas} 
\eea 
\noindent One can also explicitly check that $\vec{J} \cdot \vec{J} 
\equiv {1 \over 4} {\cal C} ({\cal C} + 2) =
{1 \over 4} \vec{a}^{\dagger} \cdot \vec{a}
(\vec{a}^{\dagger} \cdot \vec{a} + 2)$. Thus the representations of $SU(2)$
can be characterized by the eigenvalues of the total occupation number 
operator and the spin value $j$ is equal to $n/2 \equiv (n_1+ n_2)/2$.   
With the SU(2) Schwinger representation (\ref{sch}), we can directly 
generalize (\ref{cs1}) and write down the spin coherent states as:\footnote{In 
\cite{radcliffe} the SU(2) coherent state construction is through Holstein 
Primakov representation of SU(2) Lie algebra.}: 
\bea
\vert \vec{z}>_{n} & = & 
\sqrt{n!} \sum_{n_1,n_2 =0}^{n} \hspace{-0.3cm} {}^{{}^{\prime}} ~~  
\frac {z_{1}^{n_1}~~z_{2}^{n_2}}{\sqrt{n_1!~n_2!}} ~~\vert n_1, n_2>  \nonumber \\
&=& \sqrt{n!} \sum_{m =0}^{n}  ~~ \frac {z_{1}^{n-m}~~z_{2}^{m}}{\sqrt{(n-m)!~m!}} ~~\vert n-m, m>  
\label{cs2}
\eea
\noindent In (\ref{cs2}), the $\prime$ over the summation sign implies 
$n_1+n_2=n =2j$. It is easy to see that the states $|n_1,n_2 \rangle$ with 
the above constraint form $(2j+1)$ dimensional representation of SU(2)
group.  Further, $(z_1, z_2)$ is a doublet of complex numbers 
satisfying  the constraint: 
\bea  
|z_1|^2 + |z_2|^2 =1. 
\label{s3} 
\eea
Thus the coherent states in (\ref{cs2}) are defined over the sphere $S^{3}$. 
It is easy to check:
\bea
\int_{S^{3}}  d^{2}z_1~ d^{2}z_2  \delta(|z_1|^2 + |z_2|^2 -1) 
\vert \vec{z}>_{n} {}_{n}< \vec{z}| = {\cal I}_{n}  
\label{roi} 
\eea
In (\ref{roi}), ${\cal I}_{n}$ is $(n+1) \times (n+1)$ dimensional unit matrix. 
By construction, the  coherent states in (\ref{cs2}) satisfy: 
\bea 
{\cal C} \vert \vec{z}>_{n} = n ~ \vert \vec{z}>_{n}. 
\label {evc} 
\eea
Thus the  SU(2) coherent states are smoothly defined over the SU(2) group 
manifold (${\cal M} = S^{3}$) and are eigenstates of total angular momentum 
operator ($J^{2}=J_1^2+J_2^2+J_3^2$).  In terms of angular momentum basis,
$|j;m \rangle \equiv |n_1=j+m,n_2=j-m\rangle$, the spin coherent states 
can be written as: 
\bea
\vert \vec{z}>_{n=2j} & = & \sqrt{2j!}  
\sum_{m = -j}^{j} ~~  
{\frac{1}{\sqrt{(j+m)!~(j-m)!}}} 
z_{1}^{j+m}~~z_{2}^{j-m}
~~\vert j;m> 
\label{cs22}
\eea
We can also obtain (\ref{cs22}) by 
directly operating the SU(2) group element $U(\theta,\phi,\psi) \equiv \exp 
(i \phi J_3) \exp (i \theta J_2) \exp (i \psi J_3)$ on the highest weight state 
in the $j^{th}$ representation \cite{perelomov}:
\bea
\vert \theta,\phi,\psi) >_{j} ~&=&~ U(\theta,\phi,\psi) ~\vert j; j> ~,
\nonumber \\
&=&~ exp i (j\psi) \sum_{m=-j}^{+j} C_{m}(\theta,\phi) ~\vert j; m> ~,
\label{st}
\eea 
\noindent In (\ref{st}), 
\bea 
C_{m}(\theta,\phi) = \sqrt{\frac{2j!}{(j+m)!(j-m)!}}~ exp(im\phi)~ 
\Big(sin \frac{\theta}{2}\Big)^{j-m}~ 
\Big(cos \frac{\theta}{2}\Big)^{j+m}
\label{cof} 
\eea
The identification: 
\bea 
z_1 \equiv e^{i {\frac{\psi}{2}}} e^{i {\frac{\phi}{2}}} 
cos {\theta \over 2}, ~~ 
z_2 \equiv e^{i {\frac{\psi}{2}}} e^{-i {\frac{\phi}{2}}} 
sin {\theta \over 2} 
\label{eq}
\eea 
shows the equivalence of the harmonic oscillator (\ref{cs2}) and group action (\ref{st}) 
constructions. We now 
exploit Schwinger boson representation (\ref{sch}) to construct 
new types of coherent states which can not be generated by 
simple group action. 

\subsection{SU(2) Charge Coherent States} 

So far the SU(2) coherent states were defined with fixed angular 
momentum j and they were linear combinations of the states 
$|j;m \rangle$ with m varying from $- j$ to $+ j$. The corresponding 
weight factors were definite analytic functions on SU(2) manifold. 
We now define new types of SU(2) coherent states which carry fixed 
charge (m) and are linear combinations of the states $|j;m \rangle$
with j varying from $|m|$ to $\infty$. Again the corresponding weight 
factors are certain analytic functions on SU(2) manifold. 
For convenience, we define the SU(2) charge operator ${\cal Q}$ 
to be twice of the third component of the angular momentum, i.e, 
\bea 
{\cal Q}  = a_{1}^{\dagger}  a_{1} - a_{2}^{\dagger}  a_{2}.  
\label{ch}
\eea 
The SU(2) fixed charge (q=2m) coherent states are defined as: 
\bea
|\vec{z} \rangle_{q} \equiv \vert z_1, z_2>_{q} & = &  
\sum_{n_1,n_2 =0}^{\infty} 
\hspace{-0.3cm} {}^{{}^{\prime}} ~~  
\sqrt{\frac {(n_1 + n_2 +1)!}{{n_1!~n_2!}}}~z_{1}^{n_1}~~z_{2}^{n_2}~~ \vert n_1, n_2> 
\nonumber \\ 
& = & \sum_{r=0}^{\infty} \sqrt{\frac {(q+2r+1)!}{(q+r)!~r!}} 
z_{1}^{(q+r)}~~z_{2}^{r} ~~ \vert q+r, r>  
\label{cs3}
\eea
\noindent In (\ref{cs3}), the $\prime$ over $\sum$ implies that $n_1 - n_2 =q$. Therefore, 
the fixed charge coherent states in (\ref{cs3}) satisfy: 
\bea 
{\cal Q} \vert \vec{z} >_{q} & = & q ~ \vert \vec{z}>_{q}  
\label {evq} 
\eea
Further, as the total number operaotor ${\cal C}$ and $a_{1}a_2$ commute with 
${\cal Q}$, it is easy to check: 
\bea
f({\cal C}) a_1 a_2 \vert \vec{z} >_{q} & = & z_1 z_2  ~ \vert \vec{z}>_{q} 
\label{a1a2} 
\eea
where  $f({\cal C}) \equiv \frac{1}{\sqrt{({\cal C}+3)({\cal C}+2)}}$.  
We note that in the context of canonical coherent states, the states  satisfying 
$f(a^{\dagger}a) a |z> = z|z>$  are known as non-linear coherent 
states and  have been extensively studied  (see \cite{fan} and references therein).   

\noindent The charge coherent states  satisfy  resolution of identity over the SU(2) 
group manifold $S^{3}$ (see appendix): 
\bea
\int_{S^{3}} d^{2}z_1 d^{2}z_2  \delta(|z_1|^2 + |z_2|^2 -1) 
\vert \vec{z}>_{q} {}_{q}< \vec{z}| = {\cal I}_{q}  
\label{roi2} 
\eea
\noindent In (\ref{roi2}), ${\cal I}_{q}$ is the infinite dimensional 
unit matrix. 
Thus we have constructed the charge coherent states on $S^3$. 
Therefore, using resolution of identity (\ref{roi}), we can express 
them in terms of fixed angular momentum coherent states (\ref{cs2}):  
The  expansion is: 
\bea 
|z_1,z_2 \rangle_{q} = \sum_{n=0}^{\infty} \int_{S^{3}} d^2 w_1 d^2 w_2  
~~{}_{n}\langle w_1, w_2| z_1, z_2 \rangle_{q} ~~|w_1,w_2 \rangle_{n} 
\eea 
Putting $n=2j$ in (\ref{cs2}) and 
$q= 2 j_{3}$ in (\ref{cs3}), it is easy to see that  
$\vert \vec{z} >_{q=2j_3}$ has non-zero overlap with 
$\vert \vec{z} >_{n=2j}$  iff $j \geq |j_3|$.  More 
precisely: 
\bea 
{}_{n}\langle w_1, w_2| z_1, z_2 \rangle_{q} & = & {\sqrt{n!(n+1)!} \over (p+q)!p!}  
(\bar{w}_1 z_1)^{p+q}  (\bar{w}_2 z_2)^{p};  ~~~if ~~n =  2 p + q \nonumber \\
                                             & = & 0 ~~~ otherwise. 
\label{prr}
\eea
In (\ref{prr}), p is a non-negative integer such that n is positive. 
It is interesting to write (\ref{cs3}) in the angular 
momentum basis $|j;m\rangle$, 
\bea 
|\vec{z} \rangle_{q=2m} \equiv   \sum_{j=|m|}^{\infty} 
\sqrt{\frac {(2j+1)!} {(j+m)!~(j-m)!}} 
z_{1}^{(j+m)}~~z_{2}^{j-m} ~~ \vert j; m>  
\label{cs33}
\eea
Note that the fixed charge coherent state (\ref{cs33}), unlike spin 
coherent states (\ref{cs22}), can not be obtained by a simple group 
action like in (\ref{st}) and they are not eigenstates of $J^{2}$. 

At this stage it is interesting to compare our formulation of SU(2) 
charge coherent states with the already  existing formulations \cite{eriksson,bdr}. 
In \cite{bdr}, the SU(2) charge coherent states are defined as: 
\bea 
|\zeta>_{q} = N_{q} \sum_{n=0}^{\infty} {\zeta^{n} \over \sqrt{[n!(n+q)!]}}
|n+q,n> 
\label{bbdr} 
\eea
In (\ref{bbdr}) $N_{q}$ is the normalization factor and $\zeta$ are the co-ordinates 
of complex plane. Further, the above coherent states satisfy resolution of 
identity: 
\bea 
\int_{R^{2}} {d^{2} \zeta \over \pi}  ~~\phi_{q}(\zeta) |\zeta>_{q} {}_{q}<\zeta| = {\cal I}_{q} 
\label{bbsh} 
\eea
In (\ref{bbsh}),  $\phi_{q}(\zeta) = J_{q}(2i|\zeta|) K_{q}(2|\zeta|)$ where $J_{q}$,
$K_{q}$ are the Bessel and modified Bessels functions respectively. 
In \cite{eriksson}, SU(2) fixed charge 
coherent states $|z \rangle_{j,q}$ are constructed on the complex plane $z$ 
which are eigenstates of $J^{2}$ as well as the charge operator $J^{3}$. This has been possible 
because of the use of three harmonic oscillators $(a_1,a_2,a_3)$ to construct SU(2) Lie algebra: 
\bea 
\hat{J}^{i} = - i \epsilon^{ijk} a^{\dagger}_{j} a_{k}. 
\label{erik} 
\eea 
Denoting $ |n_1,n_2,n_3> = (n_1!n_2!n_3!)^{-{1 \over 2}} (a^{\dagger}_{1})^{n_1}  
(a^{\dagger}_{2})^{n_2} (a^{\dagger}_{3})^{n_3}|0,0,0> $, the SU(2) charged coherent states 
are defined as: 
\bea 
|\zeta>_{j,q} = \sum_{n=0}^{\infty} \Big(\frac{2^{j}(n+j)}{(2n+2j+1)!n!}\Big)^{1 \over 2} \zeta^{n} 
|j,q,j+2n> 
\label{erik1} 
\eea
They satisfy \cite{eriksson}: 
\bea 
J.J |\zeta>_{j,q} = j(j+1)  |\zeta>_{j,q}, ~~ {\cal Q} |\zeta>_{j,q} = q |\zeta>_{j,q}, 
~~ a.a |\zeta>_{j,q} = \zeta |\zeta>_{j,q}
\label{err} 
\eea
where  $a.a = a_1a_1+  a_2a_2 + a_3a_3$. The resolution of identity is given by: 
\bea
\sum_{j=0}^{\infty} \sum_{q=-j}^{j} \int_{R^{2}} {d^{2} \zeta \over 2 \pi} 
\Phi_{j}(|\zeta|) |\zeta>_{j,q} {}_{j,q} <\zeta| = {\cal I} 
\label{err1} 
\eea
In (\ref{err1}), $\Phi_{j}(|\zeta|)$ are related to modified Bessel functions \cite{eriksson}. 
Thus the charged coherent states $|z>_{q}$ defined in (\ref{cs3}) or (\ref{cs33}) are different 
from the ones discussed earlier in \cite{eriksson,bdr}. In particular, unlike $|z>_{q}$ which 
are defined on the SU(2) group manifold $S^{3}$, $|\zeta>_{q}$ in (\ref{bbdr}) and 
$|\zeta>_{j,q}$ in (\ref{erik1}) are defined on non-compact manifold $R^{2}$. 
In the next section, we will instead use the three oscillators 
to define  SU(3) coherent states. 

\section{\bf SU(3) Charge Coherent States}

\noindent We will now generalize the fixed SU(2) charge coherent state
ideas to the group SU(3). For the sake of simplicity, we will restrict 
ourselves to SU(3) representations which are completely symmetric. 
We, therefore,  define a  triplet of harmonic oscillator creation and
annihilation operators satisfying :
\bea 
\big[a_i,a^{\dagger}_j \big] =  \delta_{ij}~; i,j =1,2,3.   
\eea
Let $\frac{\lambda^a}{2}$, $a=1,2,...,8$ be the generators of $SU(3)$ in the 
fundamental representation; they satisfy the $SU(3)$ Lie algebra 
$[\frac{\lambda^a}{2}, \frac{\lambda^b}{2}] = 
i f^{abc} \frac{\lambda^c}{2}$. Let us define the following operators \cite{georgi}: 
\beq
Q^a =  \frac{1}{2} a^\dagger_i \lambda^a_{ij}  a_j  ~,
\label{suc} 
\eeq
More explicitly\footnote{Note that in \cite{eriksson}, $J^{1} \equiv 
2 Q^{7}, J^{2} \equiv -2 Q^{5}, J^{3} \equiv 2Q^{2}$ are used to 
define coherent states with fixed $J^{3}$ and $\vec{J}.\vec{J}$.}:  
\bea 
Q^3 ~&=&~ \frac{1}{2} ~( a_1^\dagger a_1 ~-~ a_2^\dagger a_2), ~~ 
Q^8  =  \frac{1}{2 {\sqrt 3}} ~( a_1^\dagger a_1 ~+~ a_2^\dagger a_2 ~-~ 2
a_3^\dagger a_3 ) \nonumber \\
Q^1 ~&=&~ \frac{1}{2} ~( a_1^\dagger a_2 ~+~ a_2^\dagger a_1 ), ~~ 
Q^2 ~=~ -\frac{i}{2} ~( a_1^\dagger a_2 ~-~ a_2^\dagger a_1 ) ~, \nonumber \\
Q^4 ~&=&~ \frac{1}{2} ~( a_1^\dagger a_3 ~+~ a_3^\dagger a_1), ~~ 
Q^5 ~=~ -\frac{i}{2} ~( a_1^\dagger a_3 ~-~ a_3^\dagger a_1) \nonumber \\
Q^6 ~&=&~ \frac{1}{2} ~( a_2^\dagger a_3 ~+~ a_3^\dagger a_2 ), ~~ 
Q^7 ~=~ -\frac{i}{2} ~( a_2^\dagger a_3 ~-~ a_3^\dagger a_2).
\label{su3}
\eea
\noindent It is clear that the total number operator ${\cal C} = a^{\dagger}_1 a_1 + 
a^{\dagger}_2 a_2 + a^{\dagger}_3 a_3 $  commutes with all the SU(3) generators 
in (\ref{su3}). 

\noindent   The SU(3) coherent state  analogous to (\ref{cs2}) are: 
\bea
\vert \vec{z}>_{n} \equiv \vert z_1, z_2, z_3>_{n} & = & 
\sqrt{n!} \sum_{n_1,n_2,n_3 =0}^{n} \hspace{-0.3cm} {}^{{}^{\prime}} ~~  
\frac {z_{1}^{n_1}~z_{2}^{n_2}~z_{3}^{n_3}}{\sqrt{n_1!~n_2!~n_3!}} 
~~\vert n_1, n_2, n_3>  
\label{cs4}
\eea
\noindent In (\ref{cs4}), the $\prime$ over the summation sign implies 
$n_1+n_2+n_3=n$. With this constraint, the states  $|n_1,n_2,n_3 \rangle$ form 
all the symmetric representations of SU(3).  
They are of dimensions $\frac{(n+1)(n+2)}{2}$ 
and the coherent states $\vert \vec{z}>_{n}$ satisfy: 
\bea 
{\cal C} \vert \vec{z}>_{n}   = n \vert \vec{z}>_{n}.  
\label{cassu3}
\eea
Further in (\ref{cs4}), $(z_1, z_2,z_3)$ is a triplet  of complex 
numbers satisfying  the constraint: 
\bea  
|z_1|^2 + |z_2|^2 + |z_3|^2 =1. 
\label{s5} 
\eea
Note that the SU(3) coherent states in a mixed representation of SU(3) can  be 
defined by introducing  a second independent set of oscillators $(b_1^{\dagger}, b_2^{\dagger}, b_3^{\dagger})$ 
and defining \cite{manu}: 
\bea
Q^a =  \frac{1}{2} a^\dagger_i \lambda^a_{ij}  a_j - \frac{1}{2} b^\dagger_i \lambda^a_{ji}  b_j ~,
\label{suc1} 
\eea
leading to second Casimir ${\cal C}^{\prime} = b^{\dagger}.b = b^{\dagger}_{1}b_{1}+b^{\dagger}_{2}b_{2}
+b^{\dagger}_{3}b_{3}$. 
However, in this work, to keep the discussion simple and analogous 
to SU(2) group, which has only symmetric representations, we will be interested only in the symmetric 
representations of SU(3) with a single Casimir given in (\ref{cassu3}) and ${\cal C}^{\prime}=0$.   
Now the SU(3) construction in (\ref{cs4}), (\ref{cassu3})  and the constraint (\ref{s5}) are 
analogous to the corresponding SU(2) construction in (\ref{cs2}), (\ref{evc})  and the 
constraint (\ref{s3}) respectively. Further, like in the SU(2) case it is easy to check:
\bea
\int d^{2}z_1 \int d^{2}z_2 \int d^{2}z_3  \delta(|z_1|^2 + |z_2|^2 
+ |z_3|^2-1) \vert \vec{z}>_{n} {}_{n}< \vec{z}| = {\cal I}_{n}  
\label{roi3} 
\eea
In (\ref{roi3}), ${\cal I}_{n}$ is ${1 \over 2} (n+1)(n+2)$ dimensional 
unit matrix.  Thus the coherent states in (\ref{cs4}) are defined over the 
sphere $S^{5}$. 
We now define SU(3) charge and hyper-charge operators for the symmetric representations to be: 
\bea 
{\cal Q}_1 & \equiv &  2 Q^{3} = (a^{\dagger}_1 a_{1} - a^{\dagger}_2 a_{2}) 
\nonumber \\
{\cal Q}_2 & \equiv  & 2 \sqrt{3} Q^{8} = a^{\dagger}_1a_{1} + a^{\dagger}_2
a_{2}- 2 a^{\dagger}_3 a_{3}  
\label{co} 
\eea
The  SU(3) charge, hyper-charge coherent states are given by:
\bea
|z_1, z_2, z_3 \rangle_{q,l}  =  \sum_{p=0}^{\infty} \sqrt{\frac{(3p+2l-q+2)!}{(p+l)!(p+l-q)!p!}} 
{z_1}^{p+l} {z_2}^{p+l-q }{z_3}^{p}~ |p+l, p+l-q, p \rangle 
\label{cs5}  
\eea
They satisfy:
\bea 
{\cal Q}_1 |z_1, z_2, z_3 \rangle_{q,l} = q |z_1, z_2, z_3 \rangle_{q,l}, ~~
{\cal Q}_2 |z_1, z_2, z_3 \rangle_{q,l} = (2l-q) |z_1, z_2, z_3 \rangle_{q,l}, 
\label{eve3} 
\eea
where $q=q_1$ and $2l={q_1}+{q_2}$. 
The coherent states in (\ref{cs5}) are  generalization 
of the corresponding SU(2) coherent states in (\ref{cs3})
and satisfy resolution of identity property (see Appendix).  
We can also expand (\ref{cs5}) in terms of (\ref{cs4}):
\bea 
|z_1,z_2,z_3 \rangle_{q,l} = \sum_{n=0}^{\infty} \int_{S^{5}} d^3 w_1 d^3 w_2  d^3 w_3  
~~{}_{n}\langle w_1, w_2, w_3| z_1, z_2 z_3\rangle_{q,l} ~~|w_1,w_2, w_3\rangle_{n} 
\label{112} 
\eea

The overlap is given by: 
\bea
{}_{n}\langle \vec{w}| \vec{z}\rangle_{q,l} & = &  
\frac{\sqrt{n!(n+2)!}}{(p+l)!(p+l-q)!p!} 
(\bar{w}_{1}z_{1})^{p+l} (\bar{w}_{2}z_{2})^{p+l-q}(\bar{w}_{3}z_{3})^{p} ~~ if n= 3p+2l-q
\nonumber \\
 &=& 0~~~ otherwise  
\label{113} 
\eea
In (\ref{113}), p is any non-negative integer such that n is +ve. The above expression for the 
overlap is analogous to (\ref{prr}) in the case of  SU(2). 

We now compare our SU(3) charge coherent state construction with the ones given in \cite{fan, seok}.   
These coherent states are defined as\footnote{We are following the notations of \cite{fan}. The charges 
in (\ref{cs5}) and (\ref{xxs}) are related by: $q=2\bar{q} -y$ and $l = \bar{q} + y$.}: 
\bea 
|\zeta>_{\bar{q},y} = N_{\bar{q},y} \sum_{m=0}^{\infty} \frac{\zeta^{m}}
{[m!(m+y+\bar{q})!(m+2y-\bar{q})!]^{1\over 2}} |m+y+\bar{q}, m+2y-\bar{q}, m> 
\label{xxs} 
\eea
Again, like in the SU(2) case, the SU(3) charge coherent states $|\zeta>_{\bar{q},y}$ are defined 
over $R^2$ and thus different from the ones in (\ref{cs5}) defined on the compact manifold 
$S^{5}$.

\section{SU(N) Charge Coherent States} 

\noindent We now briefly discuss SU(N) construction by  using N-plet of harmonic oscillators. 
The coherent states analogous to (\ref{cs2}) and (\ref{cs4}) are: 
\bea 
\vert z_1, z_2,...,z_N>_{n} & = & 
\sqrt{n!} \sum_{n_1,n_2,..,n_N =0}^{n} \hspace{-0.5cm} {}^{{}^{\prime}} ~~  
\frac {z_{1}^{n_1}~~z_{2}^{n_2} ... z_{N}^{n_N}}{\sqrt{n_1!~n_2!....n_N!}} 
~~\vert n_1, n_2, ....., n_N>  
\label{cs6}
\eea
In (\ref{cs6}),  $n_1+n_2+....+n_N = n$  and 
\bea 
|z_1|^{2}+|z_2|^{2}+....+|z_{N}|^{2} = 1.
\label{abbc} 
\eea 
Thus the coherent states in (\ref{cs6}) are defined on 
$S^{2N-1}$.  Note that, like in SU(3) case,  we have again restricted 
ourselves to only symmetric representations of SU(N). 
Now within the symmetric representations, the (N-1) charge 
operators can be chosen as ${\cal Q}_l = a^{\dagger}_l a_{l}- a^{\dagger}_{l+1} 
a_{l+1}$; l=1,2,...,$(N-1)$ and the corresponding eigenvalues will be denoted 
by $q_{l}$. Now we can easily generalize the charge coherent 
states in (\ref{cs3}) and (\ref{cs5}) to SU(N) group: 
\bea 
\vert z_1, z_2,...,z_N>_{q_1,q_2,..q_{N-1}} & = & 
\sum_{n_N=0}^{\infty} L \frac {z_{1}^{n_1}~~z_{2}^{n_2} ... z_{N}^{n_N}}{\sqrt{n_1!~n_2!....n_N!}} 
~~\vert n_1, n_2, ....., n_N>  
\label{cs7}
\eea
where 
\bea
L=\sqrt{(n_1+n_2+...+n_N+ (N-1))!} \nonumber \\
\eea 
and  the N occupation numbers $n_{i}= n_{N} + \sum_{j=i}^{(N-1)} q_j$ and $i=1,2,..,(N-1)$.
Like (\ref{cs6}), the charge coherent states in (\ref{cs7}) satisfy resolution of identity 
on $S^{2N-1}$. 
   
\section{Summary and Discussion} 

\noindent We have constructed SU(2) charge coherent states 
which are defined and satisfy resolution of identity over SU(2) 
group manifold $S^3$.  Further, they are eigenstates of $J^3$ 
{\it and not of} $\vec{J}.\vec{J}$.  We then defined SU(3) and 
SU(N) charge coherent states on $S^{5}$ and $S^{2N-1}$ respectively. 
The  spin coherent states have been extensively used to study the 
partition functions of SU(2) spin models  leading to useful 
semiclassical descriptions \cite{auer}. The  
SU(2) spin model  Hamiltonian is given by  
$H = \sum_{<i,j>} \vec{J}(i).\vec{J}(j)$ where $i$ and $j$ are site indices and 
$<i,j>$ denotes the nearest 
neighbours. This Hamiltonian commutes  
with $\vec{J}(i).\vec{J}(i), ~ \forall i$. Therefore, 
the the fixed angular momentum coherent state basis 
in (\ref{cs2}) has been exploited for the  
path integral formulation of the partition function \cite{auer}. 
Instead, let us consider anisotropic SU(2) 
spin models with Hamiltonians of the form $H = \sum_{<i,j>} \Big[J^{3}(i)J^{3}(j) 
+ K\big(h(i),h(j)\big)\Big]$ where $h(i) \equiv a^{\dagger}_{1}(i) a^{\dagger}_{2}(i)$
and K is a hermitian operator depending on $h(i)$ and $h(j)$. 
This Hamiltonian does not commute with $\vec{J}(i).\vec{J}(i)$  because of 
the presence of $a^{\dagger}_{1}a^{\dagger}_{2}$ and  $a_{1}a_{2}$ terms.  These terms  
change the corresponding spin value by $\pm 1$ respectively  as is clear from 
(\ref{cas}). However, these terms do not change the value of 
$J^{3} = {1 \over 2} \big(a^{\dagger}_{1} a_{1} - a^{\dagger}_{2} a_{2}\big)$.  
Therefore, to study such Hamiltonians, the fixed charge ($J^{3}$) coherent states in (\ref{cs3}) 
or (\ref{cs33}) should be useful. 

We note that the construction of SU(N) $(N \ge 3)$ charge coherent states involved only symmetric 
representations of SU(N). This was the reason we were 
led to a simple SU(N) generalization of SU(2) results.  
It would be interesting to consider all the irreducible representations of SU(N) 
and define the most general SU(N) charge coherent states. This can 
be done by including $(N-1)$ sets of harmonic oscillators 
belonging to all the $(N-1)$ fundamental representations of SU(N) \cite{manu}. 
The work in this direction is in progress and will be reported elsewhere. 

\begin{flushleft} 
{\bf Appendix} 
\end{flushleft} 

\noindent In this appendix we prove that the SU(3) charge, 
hyper-charge coherent states in 
(\ref{cs5}) satisfy resolution of identity property. 
\bea 
{\cal I}_{q,l} = \int d^2 z_1  d^2 z_2  d^2 z_3 \delta(|z_1|^2 + |z_2|^2 + |z_3|^2 -1) 
|z_1,z_2,z_3 \rangle_{q,l}~{}_{q,l} \langle z_1,z_2,z_3|
\label{114}
\eea
To solve the $\delta$ function constraints, it is convenient to define
\bea
z_1 = r_1 e^{i\theta_1}, ~~~z_2 = r_2 e^{i\theta_2}, ~~~z_3 = r_3 e^{i\theta_3}. 
\nonumber
\eea 
After integrating out the angles $\theta_1,\theta_2$ and $\theta_3$, we 
get: 
\bea
{\cal I}_{q,l}  =  C \sum_{n=0}^{\infty} \frac{(3n+2l-q+2)!}{(n+l)!(n+l-q)!n!} 
\int r_1 dr_1  r_2 dr_2 r_3 dr_3 \delta(r_1^{2}+r_2^{2} +r_3^{2}-1) 
(r_1^2)^{n+l} (r_2^2)^{n+l-q} 
\nonumber \\
~~~~~(r_3^2)^{n}|n+l,n+l-q,n \rangle \langle n+l,n+l-q,n|
\label{115} ~~~~~~~~~~  
\eea  
In (\ref{115}), C is a constant. 
In terms of the above radial co-ordinates the $\delta$ function constraint
can be solved by polar decomposition, i.e; 
\bea
r_1 = r sin\theta cos\phi, ~~~r_2 = r sin\theta sin\phi, ~~~ r_3 = r cos\theta
\label{116} 
\eea
In (\ref{116}), since $r_1,r_2,r_3 \ge 0$, both $\theta$ and $\phi$ vary 
between 0 and $\frac{\pi}{2}$. Now r can be trivially integrated off because 
of the $\delta$ function and we are left with: 
\bea 
{\cal I}_{q,l}  =  C \sum_{n=0}^{\infty} \frac{(3n+2l-q+2)!}{(n+l)!(n+l-q)!n!} 
\int_{0}^{\frac{\pi}{2}} sin\theta cos\theta d\theta 
(sin^2\theta)^{2n+2l-q+1} (cos^2\theta)^{n}  \nonumber \\ 
\int_{0}^{\frac{\pi}{2}} 
sin\phi cos\phi d\phi  (sin^2\phi)^{n+l-q} (cos^2\phi)^{n+l} 
|n+l,n+l-q,n \rangle \langle n+l,n+l-q,n|
\label{117}  
\eea  
We now set: $ x=sin^2\theta, ~~~ y =sin^2\phi$ to get 
\bea 
{\cal I}_{q,l} = C \sum_{n=0}^{\infty} \frac{(3n+2l-q+2)!}{(n+l)!(n+l-q)!n!} 
            \int_{0}^{1} dx x^{2n+2l-q+1} (1-x)^{n} 
            \int_{0}^{1} dy y^{n+l-q} (1-y)^{n+l} \nonumber \\ 
|n+l,n+l-q,n \rangle \langle n+l,n+l-q,n| \nonumber \\
\label{118} 
\eea
Using the fact: 
\bea
B(m,n) = \int_{0}^{1} dx x^{m-1} (1-x)^{n-1}= \frac{(m-1)!(n-1)!}{(m+n-1)!} \nonumber 
\eea
we get: 
\bea
{\cal I}_{q,l} = C \sum_{n=0}^{\infty}  
|n+l,n+l-q,n \rangle \langle n+l,n+l-q,n| \nonumber 
\eea
Therefore, ${\cal I}_{q,l}$ is an identity operator in the Fock 
space of 3 harmonic oscillators with fixed SU(3) charge and hyper-charge. 
In the full Hilbert space of 3 harmonic oscillators, ${\cal I}_{q,l}$ 
is the  projection operator which projects out the fixed charge component characterized 
by q and l.  The above proof of resolution of identity can be easily generalized to
SU(N) by using polar decomposition analogous to (\ref{116}) on $S^{2N-1}$.

\end{document}